\documentclass[prd,aps,twocolumn,tighten]{revtex4}
\usepackage{latexsym}
\begin{document}
\newcommand{\beq}{\begin{equation}}
\newcommand{\eeq}{\end{equation}}
\newcommand{\Prd}{Phys. Rev D}
\newcommand{\Prl}{Phys. Rev. Lett.}
\newcommand{\Plb}{Phys. Lett. B}
\newcommand{\Cqg}{Class. Quantum Grav.}
\newcommand{\Np}{Nuc. Phys.}

\title{The cosmological dependence of weak interactions}
\author{M.Novello and P Rotelli}
\address{\mbox{}\\
International Centre for Theoretical Physics,\\
Miramare, PO Box 586, 34100 Trieste, Italia\\
E-mail: novello@cbpf.br}
\date{15 October 1972}

\begin{abstract}
A model for the cosmological time dependence of weak interactions
is discused and some experimental tests suggested.
 \end{abstract}

\maketitle
\newpage

\section{Introduction}

In this paper a model is described which suggests a link between
gravitational and weak interactions. Thus any time dependence in
the gravitational interaction appears also in the weak
interactions. In $\S$ 2 the model is presented and the specific
form that the time variation of the weak interactions takes is
developed. In $\S$ 3 some of the consequences of the model
involving laboratory neutrinos are discussed. Section 4 describes
a possible program for testing the model by the detection of
cosmic neutrinos. We conclude with $\S$ 5 in which some further
speculations about the time dependence of physical laws are made.

\section{The cosmological model}

The idea that interactions may change with time stems from a paper by
Dirac (1937) in which the gravitational constant was treated as time
dependent. However, it was later realized that, as originally
expressed, his hypothesis contradicted the principle of covariance.
Jordan (1961, {\it Problems in Gravitation unpublished}) and others (Dicke
1963) have since produced a way of circumventing this difficulty by
introducing a scalar field $\phi (x)$ into the theory. Recently this
$\phi$
field has been used to develop singularity-free cosmological models
(Novello 1971, unpublished) by employing a nonlinear Lagrangian in a
scalar-tensor theory of gravitation. To obtain this well-behaved form
of the Universe the $\phi$ field must have a regular minimum at $t=0$
\begin{equation}
\phi (t)\sim \phi_0 + \phi_1 t^2 +\cdots \quad \mbox{(for small $t$)}
\end{equation}
and go to a constant $Q$ for large values of $t$
\begin{equation}
\phi
(t){\longrightarrow}_{{{}_{}}_{\!\!\!\!\!\!\!\!\!\!\!\!\!\!\!t\rightarrow
\infty}} \ Q \ .
\end{equation}

Now, the riemannian structure of space-time implies that the
generalized $\gamma$, which define the metric tensor by the
anticommutation relation
\begin{equation}
\left\{\gamma_\alpha(x),\ \gamma_{\beta}(x)\right\}=2g_{\alpha
\beta}(x)1\!\!1 ,
\end{equation}
obey the equation
\begin{equation}
\gamma_{\alpha||\beta}(x)=\sigma
\left[U_{\beta}(x),\gamma_\alpha(x)\right]
\end{equation}
where $\sigma$ is a constant,
$U_{\beta}(x)=\gamma_{\beta}(x)(1\!\!1+ \gamma_5(x))$ and double
bar ($||$) means covariant derivative, that is
\[
\gamma_{\alpha||\beta}(x)=\gamma_{\alpha|\beta}(x)-\Gamma^{\varepsilon}_{\alpha\beta}(x)
\gamma_{\varepsilon}(x)+ \left[\tau_{\beta}(x),\gamma_\alpha
(x)\right]
\]
where the single bar means the usual derivative,
$\Gamma^{\varepsilon}_{\alpha \beta}$  are the connections of
the Riemann space and $\tau_{\alpha}$  are {\it internal} connections
that arise from the permissible generalized gauge transformation
\[
\gamma_{\alpha}(x)\rightarrow
\gamma'_\alpha(x)=M(x)\gamma_\alpha(x)M^{-1}(x)
\]
for an arbitrary matrix $M(x)$.

Equation (4) is the most general expression consistent with the
riemannian structure of the space that can be constructed with the
elements of the Clifford algebra without including any arbitrary
extra field.

It has previously been shown (Novello 1971) that starting from
this evolution operator $U_\lambda$, (equation (4)) for the
generalized Clifford algebra (space-time dependent) one can arrive
at a modified class of Einstein's equations that relate the
riemannian contracted curvature tensor with $I\!\!R_{\alpha
\beta}$ the curvature of the internal space
\begin{equation}
R_{\alpha \beta}(x)\gamma^{\alpha}(x)+
\left[I\!\!R_{\alpha \beta}(x),\gamma^{\alpha}(x)\right]=0\ .
\end{equation}
This immediately suggests a link between gravitation and weak
interactions because, given the form of $U_\lambda(x)$, the only
nontrivial interaction Lagrangian which can be constructed from
$U_\lambda$ and spinor fields is the current-current interaction
\begin{equation}
{\cal L}_1= \frac{G}{\sqrt{2}}\ J_{\beta}J^{\beta}
\end{equation}
where
\begin{equation}
J_{\beta}(x) = \bar{\psi}(x)\gamma_{\beta}(x)(1\!\!1+\gamma_5
(x)\psi (x)\ .
\end{equation}
The above consideration induces us to propose that the modified form
of Dirac's idea should be applied not only with respect to the
gravitational interaction but also to the weak interactions. In a
homogeneous and isotropic cosmological model--such as the one we are
considering--we shall see that the influence of cosmology on the
weak interactions produces a time-dependent weighting of the axial
vector current relative to to the vector current. A direct way to do
this is to consider the Lagrangian (6) and compare it with the usual
flat-space Lagrangian
\begin{equation}
{\cal L}'=\frac{G}{\sqrt{2}}\ j_{\alpha}j^{\alpha}
\end{equation}
where
\begin{equation}
j_\alpha=\bar{\psi}\gamma_{\alpha} \,(1+\gamma_5)\psi
\end{equation}
($\gamma_{\alpha}$, and $\gamma_5$ being constant Dirac matrices)
In the particular type of universe we are considering we may write
\begin{equation}
\gamma_\alpha(x)=F(\alpha,x)\gamma_\alpha \ .
\end{equation}
Indeed, from (3) and because in the co-moving system of coordinates
\begin{equation}
ds^2=dt^2-F_1(dx^1)^2-F_2(dx^2)^2-F_3(dx^2)^2
\end{equation}
it follows that
\begin{eqnarray}
\gamma_0(x)&=&\gamma_0\nonumber \\
\gamma_1(x)&=&F^{1/2}_1\gamma_1\nonumber \\
\gamma_2(x)&=&F^{1/2}_2\gamma_2\nonumber \\
\gamma_3(x)&=&F^{1/2}_3\gamma_3\ .\nonumber \\
\end{eqnarray}
Substituting (12) and (11) into (7) yields for the interaction
defined by (6) the  expression
\begin{eqnarray}
{\cal L}_1 &=& \frac{G}{\sqrt{2}}
\,\Big\{\big\{\bar{\psi}\gamma_0(1+\varepsilon (x)\psi\big\}
\big\{\bar{\psi}\gamma^0(1+\varepsilon
(x)\gamma_5\psi\big\}\nonumber
\\
&+&\big\{\bar{\psi}\sqrt{F_1}\gamma_1(1+\varepsilon
(x)\gamma_5)\psi\big\}\nonumber \\
&& \Big(\bar{\psi}\ \frac{1}{\sqrt{F_1}}\ \gamma^1(1+\varepsilon
(x) \gamma_5)\psi \Big)+\cdots \Big\}\nonumber
\end{eqnarray}
which shows that the only effective modification of generalizing to
the $\gamma_{\alpha}(x)$ functions is a space-time weighting of the
axial vector
current relative to the vector current (the vector current
modification being absorbed in ${\cal L}_1$ by the modified space-time metric
tensor). Thus we may write the weak leptonic current as
\begin{equation}
J_{\alpha}=\bar{\psi}\gamma_{\alpha}(1+\varepsilon (x)\gamma_5)\psi
\end{equation}
where $\varepsilon (x)$ is a function of $\phi (x)$ and the simplest
assumption would be that they are linearly related, that is
\begin{equation}
\varepsilon (x)=\frac{1}{Q}\ \phi (x)
\end{equation}
whence the maximal violation of parity in the $V-A$ theory is reached
only at asymptotic cosmological time.

\section{Consequences for the weak interactions}

The model has two obvious consequences: (i) Since we do not exist at
asymptotic cosmological time, the present leptonic weak current does
not violate parity maximally. (ii) Produced neutrinos and
antineutrinos are admixtures of both {\it left} and {\it right}
polarized states.
The ratio of the admixture depends upon their (cosmological) time of
creation.

The first consequence can best be tested by a very accurate
laboratory measurement of the Michel parameter $\rho$ in $\mu$
meson decay. Let us define a parameter $\delta$ by writing the
present weak leptonic current as
\begin{equation}
J_\alpha=\bar{\psi}\gamma_{\alpha}\left\{1 + (1
-\delta)\gamma_5\right\}\psi
\end{equation}
that is, $\varepsilon (t_0)=1-\delta$ where $t_{0}$ is our present
cosmological time. The $V-A$ theory\footnote{We of course employ
the lepton number-conserving charged currents. Often, however,
(and particularly in $\mu$ decay) $V,\ A, \ S,\ T$ and $P$ are
defined for the `charge retention' currents. Our modification of
the $V-A$ theory corresponds to the appearance of $S-P$ terms in
addition to $V-A$ in the charge retention current.} appears in the
limit $\delta =0$. If we then neglect, for the moment, all masses
involved except for the $\mu$ meson mass and neglect the
calculable radiative corrections, we find that
\begin{equation}
\rho = \frac{3\left(1-2\delta +\frac{3}{2}\ \delta^2\right)}
{4\left(1-2\delta +2\delta^2\right)}\simeq \frac{3}{4}
\left(1-\delta^2/2\right)\ .
\end{equation}
Since, as we shall see, this may be a very small modification to
the usual value of $\frac{3}{4}$, we have recalculated the decay
rated $W$ for a polarized muon retaining the electron mass and the
$\mu$ neutrino mass $(< 1.15 \ MeV)$ but neglecting the electron
antineutrino mass because of its very low experimental upper limit
$(< 60\ eV)$. We  find to order $\delta^2$
\begin{eqnarray}
\frac{dW}{dEd\cos \theta}&=& \frac{(1-\gamma )^2G^2\left(1-2\delta
+\frac{1}{2}\ \delta^2\right)|k|} {24\pi^3}\nonumber
\\
&& \Big(E\left(\mu^2+e^2-2\mu E\right)\left(1+3\delta^2-\gamma
\right) \nonumber \\
&+& 2(\mu -E)\left(\mu E-e^2\right)\nonumber \\
&& (1+2\gamma )-3\nu \left(\mu
E-e^2\right)\delta (1+\delta /2)+\frac{\alpha}{2\pi}f(E)\nonumber \\
&+& |k| \cos \theta  (\mu^2+3e^2-2\mu E) \nonumber \\
 &-& |k| \cos \theta \left( \gamma (\mu^2-3e^2+2\mu E)+ \frac{\alpha}{2\pi} g(E)\right)\Big)
\end{eqnarray}
where $E, \ |k|$ and e are the electron energy, magnitude of
momentum $\left(|k|=(E^2-e^2)^{\frac{1}{2}}\right)$ and mass,
respectively, while $\mu$ and $\nu$ are the masses of the muon and
muon neutrino, respectively. $\gamma =\nu^2/\left(\mu^2+e^2-2\mu
E\right)$ and, since the kinematically allowed values for the
electron energy run from $E_{\min}=e$ to
$E_{\max}=\left(\mu^2+e^2-\nu^2\right)/2\mu$, it follows that the
term $(1-\gamma )^2$ vanishes as $E\rightarrow E_{\max}$. Thus, as
is well known, an accurate determination of the electron energy
spectrum will yield at least an upper limit for the muon neutrino
mass. The functions $f(E)$ and $g(E)$ represent the effects of
radiative corrections. As a first approximation, and in order to
obtain an upper limit on $\delta$, we may use the first-order
corrections calculated by Kinoshita and Sirlin (1959).
Experimentally no disagreement with the $V-A$ theory has yet been
found. Indeed, allowing for the above radiative corrections
(Bardon {\it et al} 1965, Derenzo 1969)
\begin{equation}
\rho_{\exp}=0.75\left(
\begin{array}{c}
1\\
2\\
\end{array}
\right)\pm 0.003 \ .
\end{equation}
If we allow ourselves up to 1 STD we can set an upper limit to $\delta$ of
\begin{equation}
\delta < 0.05 \ .
\end{equation}
The electromagnetic corrections to $\mu$ decay are particularly
important, introducing an effective diminution of several per cent
in $\rho$. Thus to determine $\delta$ we shall require not only
more accurate experiments but also theoretical calculations of
second-order radiative corrections\footnote{It should also be
noted that these second-order corrections will require a cut-off
to cope with the ultraviolet divergences as indeed do the
first-order radiative corrections if note is taken of the
deviation from an exact $V-A$ model.} (Marshak et al 1969).

A similar determination of an upper limit to $\delta$ follows from the
modification in our model from the expression for the polarization
$(P_e)$ of the produced electron in $\beta$ decay. Following the usual
assumption that the nucleons in this process are effectively at rest,
and integrating over the outcoming antineutrino three-momentum, we find that
\begin{eqnarray}
dW &\propto&
 (1+3\eta^3) m^2E_eE_{\bar{\nu}} [(1-\zeta )\big\{
\left(1-\delta /2\right)^2 \,\left(1+v_e\right) \nonumber \\
&+& \delta^2\left(1-v_e\right)/4\big\} +(1+\zeta )\big (1-\delta
/2)^2(1-v_e)\nonumber \\
&+& \delta^2\left(1+v_e\right)/4\big\}]
\end{eqnarray}
where $\eta =g_A/g_v$ for the hadronic current and where $\zeta$ is
the electron polarization vector, $v_e$ its relativistic velocity and
$E_e$ and $E_{\bar{\nu}}$ the energies of electron and antineutrino,
respectively. Thus we find that
\begin{equation}
P_e=\frac{R-L}{R + L}=-v_e\left(1-\delta^2/2\right)\ .
\end{equation}
Experimentally (Willis and Thompson 1968)
\begin{equation}
P_{e^-}/v_{e^-}=-1.001\pm 0.008
\end{equation}
which, allowing for 1 STD, sets an upper limit to $\delta$ of
\begin{equation}
\delta < 0.12
\end{equation}
Thus we conclude that although no direct evidence for a nonzero
$\delta$ exists, the above data only set an upper limit of about
one-twentieth on its value.

\section{Cosmic neutrinos}

As a test of the second consequence listed in $\S$ 3 we first note that
a neutrino (antineutrino) produced at cosmological time $t$ by a weak
current of the form $\bar{\psi}\gamma_{\alpha}\left(1+\varepsilon (t)
\gamma_5\right)\psi$ can be written as
\begin{equation}
\psi^{(t)}_{\nu (\bar{\nu})}=\cos \theta (t)\psi^{L(R)}_{\nu (\bar{\nu})}+
\sin \theta (t)\psi^{R(L)}_{\nu (\bar{\nu})}
\end{equation}
where by $L$ and $R$ we mean left and right polarizations,
respectively, and where tan $\theta (t)=\left(1-\varepsilon
(t)\right) /\left(1+\varepsilon (t)\right)$. If, as is usually
assumed, we lived in a world in which $\delta =0$ and the
neutrinos were massless, then the right-(left-)handed polarized
neutrinos (antineutrinos) would be completely invisible to any
detection apparatus (save possibly one employing the gravitation
interaction). Thus the only observable consequences of the model,
for the detection of cosmic neutrinos (antineutrinos) on the
Earth, would be the effective diminution of the universal Fermi
constant $G$ by $\cos \theta (t)$. This could be measured in the
{\it Gedenken} experiment in which all $L$ neutrinos ($R$
antineutrinos) from a particular source are absorbed and counted
during a specified time. This provides us with an experimental
measurement of the otherwise elusive flux, and together with a
measurement of their rate of interaction the `effective' Fermi
constant could be deduced. Of course this is impossible in
practice because of the very low interaction rate of the neutrinos
(antineutrinos), and very few of those entering a laboratory will
be detected. It is this very property, however, which in part
makes cosmic neutrinos so interesting, for if detected they may
well carry information from very distant sources.

A feasible experiment depends upon a nonzero $\delta$, since this
will allow $N$ us, in principle, to measure both $\psi^{L(R)}_{\nu
(\bar{\nu}}$ and $\psi^{R(L)}_{\nu (\bar{\nu}}$. Indeed the rates
of interaction of these components, from a particular source, are
proportional to $\cos^2 \theta (1-\delta /2)^2 \times$ flux and
$\sin^2 \theta (\delta^2/4)\times$ flux, respectively (ignoring
for simplicity any possible mass for the neutrino). Since the flux
is the same in both cases, the ratio of these rates determines
$\delta^2 \tan^2 \theta$. This ratio appears, for example, in the
polarization of electrons produced via inverse $\beta$ decay (for
low momentum transfer):
\begin{equation}
\frac{P_e}{v_e}\simeq \frac{\delta^2\tan^2 \theta -4}
{\delta^2\tan^2\theta +4}\ .
\end{equation}

It is therefore conceivable that some cosmic neutrinos produce
unpolarized electrons. To proceed further and try to determine the
time dependence of $\theta (t)$ we would require the development of a
`neutrino telescope', that is, an equipment in which the momentum of
the incoming neutrino (antineutrino) could be deduced from the
outcoming particles of its interaction. It is probable that neutrinos
(antineutrinos) from a specific direction are dominated by one source
and that the cosmological age of the source will vary with direction.
Thus a measurement of the ratio in equation (25) would also vary with
direction indicating a time dependence of $\theta (t)$. If in addition these
sources could be identified with known radio or optical sources the
actual dependence on time of $\theta (t)$ could be found. Such a `telescope'
could also be used to determine accurately any mass of the neutrino
by measuring the time delay between neutrinos and photons produced in
for example, an exploding star, as has already been suggested
(Pontecorvo 1968).

Unfortunately this program, in spite of some optimistic plans
(Ginsburg 1971) is hampered by the low energies envisaged (due to the
Doppler effect) for cosmic neutrinos, their predicted low flux and
the failure up to the present time of attempts to measure the much
more plentiful neutrinos expected from the sun.

\section{Conclusions}

It follows from what has been said in the previous section that the
test of the model by direct measurement of cosmic neutrinos is not
likely in the near future. There is, however, another consequence of
the model which may prove relevant, and that is that because we now
predict the existence of a full complement of four neutrinos (and
four antineutrinos) the usual estimates of the maximum energy density
of the neutrino sea in the Universe become doubled.

It is tempting both from the upper limits set on $\delta$ and by
the very `anaesthetic' nature of its consequence (the breakdown in
the maximal violation of parity) to suggest a link between this
effect and CP violation. In this vein CP violation would, like
$\delta$ itself, be a vanishing phenomenon, and in addition one
would expect the two phenomena to be of the same order of
magnitude, that is, about 10$^{-3}$ smaller than the weak
interaction in general. If this is the case then present
experimental accuracies, particularly in $\mu$ decay, are not far
short of the mark.

We expect no modification in the strength of the electromagnetic
charge coupling because of the CVC (conserved vector current)
hypothesis relating it to the unmodified vector part of the weak
leptonic current. But we might have a parity-violating term of the
kind $J_{5\mu}A^{\mu}$ which, in analogy with the CP-violating term in weak
interactions, is a vanishing function with cosmological time.

Finally, since we have discussed. the possible time variation in
gravitational, weak and even electromagnetic interactions, all linked
to the time dependence of the scalar $\phi$ field, it would be unjust not
to contemplate the possible time dependence of the strong
interactions themselves (Davies 1972). In general we are advocating a
theory in which the physical laws are a function of space time
(appearing constant only locally).
It may in fact well be that the existence of `unexplained' energy
sources in the Universe is just a consequence of our microscopic view
of the laws of physics.

\section*{Acknowledgments}

One of us (MN) would like to thank Dr M A Gregorio for useful
discussions. We are grateful to Professors Abdus Salam and P Budini
as well as the International Atomic Energy Agency and UNESCO for
hospitality at the International Centre for Theoretical Physics,
Trieste.

\section*{References}

\begin{description}
\item
[]Bardon M {\it et al} 1965 {\it Phys. Rev. Lett.} {\bf 14} 449-53
\item
[]Davies P C W 1972 {\it J. Phys. A: Gen. Phys.} {\bf 5} 1296-304
\item
[]Derenzo S E 1969 {\it Phys. Rev.} {\bf} 181 1854-66
\item
[]Dicke R H 1963 {\it Relativity, Groups and Topology}
(New York: Gordon and Breach) pp 165-315
\item
[]Dirac P A M 1937 {\it Nature, Lond.} {\bf 139} 323
\item
[]Ginsburg V L 1971{\it Sov. Phys.-Usp.} {\bf 14} 21-39
\item
[]Kinoshita T and Sirlin A 1959 {\it Phys. Rev.} {\bf 113} 1652-60
\item
[]Marshak R E, Riazuddin and Ryan C P 1969
{\it Theory of Weak Interactions} (New York: Wiley)
\item
[]Novello M 1971 {|it J. math. Phys.} {\bf 12} 1039-41
\item
[]Pontecorvo B 1968 {\it Sov. Phys.-JETP} {\bf 26} 984-8
\item
[]Willis W J and Thompson J 1968 {\it Advances in Particle Physics,}
eds R L Cool and R E Marshak (New York: Wiley) pp 295-477
\end{description}
\end{document}